\documentclass[print]{aa}
\usepackage{txfonts,graphicx,natbib,epsfig,sidecap}
\usepackage{rotating}
\usepackage{fancybox,amssymb,graphics,subfigure}

\def\Mesz{M\'esz\'aros~}

\def\Bj{Bj$\ddot{\rm{o}}$rnsson~}
\def\Jo{J\'ohannesson~}

\newcommand{\beq}{\begin{equation}}
\newcommand{\eeq}{\end{equation}}
\newcommand{\ba}{\begin{eqnarray}}
\newcommand{\ea}{\end{eqnarray}}

\titlerunning{Hints of precession of central engine}

\begin{document}
\title{GRB 060206: hints of precession of the central engine?}
\author{X.W. Liu\inst{1,2,3}, X.F. Wu\inst{1,2,3,4}, T. Lu\inst{1,2,3}}

\institute{Purple Mountain Observatory, Chinese Academy of
  Sciences, Nanjing 210008, China \and National
  Astronomical Observatories, Chinese Academy of Sciences, Beijing,
  100012, China \and Joint Center for Particle Nuclear
  Physics and Cosmology of Purple Mountain Observatory - Nanjing
  University, Nanjing 210008, China \and Theoretical
  Astrophysics 130-33, California Institute of Technology, Pasadena,
  California 91125 }\mail{X.W. Liu (xwliu@pmo.ac.cn),
  X.F. Wu (xfwu@caltech.edu), T. Lu (t.lu@pmo.ac.cn)}

\date{Received / Accepted}

\abstract{}{The high-redshift ($z=4.048)$ gamma-ray burst
  GRB 060206 showed unusual behavior, with a significant rebrightening by
  a factor of $\sim4$ at about 3000 s after the burst. We
  argue that this rebrightening implies
  that the central engine became active again after
  the main burst produced by the first ejecta, then drove
  another more collimated jet-like ejecta with a larger viewing angle. The
  two ejecta both interacted with the ambient
  medium, giving rise to forward shocks that propagated into the
  ambient medium and reverse shocks that penetrated into the
  ejecta. The total emission was a combination of the emissions from
  the reverse- and forward- shocked regions. We discuss
  how this combined emission accounts for the observed rebrightening.}
{We apply numerical models to calculate the light curves from the
  shocked regions, which include a forward shock originating in the
  first ejecta and a forward-reverse shock for the second ejecta.}
{We find evidence that the central engine became
active
  again 2000 s after
  the main burst. The
  combined emission produced by interactions of these two
    ejecta with the ambient medium can describe the
    properties of the afterglow of this
    burst.
  We argue that the rapid rise in brightness at $\sim3000$
  s in the afterglow is due to the off-axis emission from the second
  ejecta. The precession of the torus or accretion
  disk of the central
  engine is a natural explanation for the departure of the second
  ejecta from the line of sight.}{}


\maketitle

\section{Introduction}
The gamma-ray burst (GRB) 060206 at Galactic Coordinates $l=78.07$
deg, $b=78.28$ deg triggered $Swift$-\textbf{BAT} on February 6th, 04:46:53
UT (trigger time $t=0$) (Morris et al. 2006). It
exhibited a single peak, with a duration of ${T_{90}=7\pm
2}$ s and a total fluence of $8.4\pm0.4\times10^{-7}$
$\rm{erg/cm^2}$ in the 15-350 keV band (Palmer et al. 2006). The
spectroscopic redshift $z$ is 4.048 (Fynbo et al. 2006). Applying the peak energy
${E_{\rm{peak}}}={75.4\pm19.5}$ keV, the best-fit low energy photon index
${\Gamma_1=1.06\pm 0.34}$ and a fixed high energy
photon index $\Gamma_2=2.5$, the isotropic-equivalent energy
integrated from 1 to $10^4$ keV in the explosion rest frame is
${E_{\gamma,\rm{iso}}=5.8\times10^{52}}$ erg (Palmer et al. 2006).

$Swift$-$\rm{XRT}$ began to observe this burst 58 s after the \textbf{BAT}
trigger time. At the same time, $Swift$-UVOT started the on-target
monitoring and detected the optical afterglow (Boyd et al. 2006).
A number of ground-based telescopes performed follow up
observations. The 2-m robotic Liverpool Telescope began to observe
it at $t=309$ s and carried out multicolor $r'i'z'$ photometry. In
the R-band the light-curve exhibited three obvious bumps in the
first 75 minutes including a steep rise ($\Delta r'\approx -1.6$
at $t\approx 3000$ s) (Monfardini et al. 2006). About 48.1 minutes
later after the trigger time, the Rapid Telescopes for Optical
Response (RAPTOR) system at Los Alamos National Laboratory began
to take optical images. The obtained light curve
confirmed the rebrightening from ${r'\sim 17.3}$ to a peak value
$r'\sim 16.4$. The subsequent decline to $r'{\sim 16.75}$ at
$t=80$ min was followed by a secondary rebrightening by
$\Delta r'\sim -0.1$ around $t=90$ min ($\rm{Wo\acute{z}niak}$ et
al. 2006). The MDM telescope observed a smooth break at $t_b=0.6$
days with another bump at $t\approx16000$ s. The overall X-ray
light curve has a similar shape as the optical light curve (Stanek
et al. 2007).

One of the most remarkable features of this burst is that the
optical light curve had a significant rebrightening and exhibited small
``bumps'' and ``wiggles''. Similar bumps and wiggles have also
been seen in a number of optical afterglows (Stanek et al. 2007).
GRB 970508 had an optical afterglow light curve rather
similar to that of GRB 060206 (Galama et al. 1998). The
optical light curve of another recent burst, GRB 060210, also
displayed a rebrightening at time $t\sim500$ s and a
shallow decay in the early epoch. The above ``unusual'' behavior
, which is not predicted by the standard fireball model, may be more the norm
than the exception (Stanek et al. 2007).

Possible scenarios for the remarkable rebrightening in GRB 060206
at $\sim$3000 s are a renewed energy injection (Rees \& \Mesz
1998; Kumar \& Piran 2000; Sari \& \Mesz 2000) and a density-jump
in the circum-burst medium (Dai \& Lu 2002). However, as discussed
by Monfardini et al. (2006), the X-ray band frequency is above the
cooling frequency at ${t\sim 3000}$ s, so the flux does not depend
on the ambient density (Freedman \& Waxman 2001). Nakar
\& Granot (2007) showed that even a sharp and large increase
in the ambient medium density cannot produce a significant
rebrightening as seen in the afterglow. So the rebrightening
cannot be due to a density jump in the ambient medium. If the
rebrightening is caused by energy injection, a huge impulsive
energy injection ${\Delta E\sim 1.8 E_0}$ at $\sim$3000 s was
required, where $E_0$ is the blast wave energy before the
rebrightening.

In this paper, we present an alternative scenario. The central
engine of this burst become active again after the
initial burst and ejects another more collimated jet
with a larger viewing angle. This jet and the initial jet sweep up
the interstellar medium (ISM). The multi-wavelength emission
predicted by this model
can reproduce both the observed X-ray and optical data.
The observational results are presented in Sect 2. We describe the
scenario in Sect 3 and fit the remarkable optical
rebrightening of GRB 060206 in Sect 4. Finally, we summarize our
results and discuss their implications in Sect 5.

\section{Observations}
The R-band light curve of GRB 060206 contains five distinct bumps
detected by telescopes mentioned in Sect 1. Monfardini et al. (2006)
attempted to describe the optical afterglow by applying an empirical
model, which consists of a sum of smoothly-connected broken
power-law functions \beq F_{\nu}(t)=\sum_j
F_j[\frac{2}{(t/t_j)^{-\alpha_{1,j}\cdot
n}+(t/t_j)^{-\alpha_{2,j}\cdot n}}]^{1/n}, \eeq where $\alpha_{1,j}$
and $\alpha_{2,j}$ are the pre-bump and post-bump temporal indices
of the $j$th-bump, and $n$ is the sharpness parameter. By fitting
the R-band light curve, they are able to obtain temporal information
about every bump and the break at 53000 s of post-break index
$\alpha=1.79\pm0.11$ (see Table 2 of Monfardini et al. 2006). The
post-bump indices of the 1st, 2nd, and 4th bumps are
$\alpha_2=$$1.2\pm 0.5$, $\sim1.0$ and $0.95\pm0.02$, respectively,
which are consistent with a typical value $3(p-1)/4$ or $(3p-2)/4$
in the normal decay phase predicted by the standard afterglow model
(Sari et al. 1998). If we fit the 5th bump independently, its
post-bump index follows obviously the normal decay slope. As shown
in Stanek et al. (2007), the X-ray light curve demonstrated similar
behavior to the optical light curve. There were only two X-ray
observations before $\sim 3000$ s of temporal index $\alpha=1$. The
later X-ray light curve exhibited clear short timescale variations,
but corresponding bumps were not seen.

Multi-epoch spectral energy distribution (SED) analysis revealed
either an SED evolution or an additional unresolved activity (or
both) during the early time interval from $t\sim1000$ s to
$t\sim3000$ s based on the $i'$ and $z'$ photometric data
(Monfardini et al. 2006). At a later time, the infrared to X-ray
fluxes (after a significant rebrightening) can be fitted by a single
power law with a spectral index $\beta=0.93\pm 0.02$. However, a
broken power law with $\beta_{OPT}=0.7$ and $\beta_X=1.2$ cannot be
ruled out.

\section{Scenario}
It is generally accepted that long GRBs (duration$>$2 s)
originate in collapsars (Woosley 1993;
MacFadyen \& Woosley 1999), while short GRBs,  of duration
less than 2 s, are associated with the merging of compact objects
(see Nakar 2007 for a review). In both scenarios of GRB
origins, a hot and dense accretion disk formed possibly around the GRB central engine.
Reynoso et al. (2006) argued that a rotating black hole
could induce the surrounding neutrino-cooled accretion disk
to precess and nutate. The precession period varied between
approximately $0.01$ s and $10$ s for typical central engines.
Therefore, even if the jet producing the main burst is along
the line-of-sight (LOS), the collimated outflow supplied
by the late activity of the central engine is possibly
off-LOS due to the precession of either the torus or the accretion
disk. The interaction of the off-LOS outflow with
the ambient medium could play an important role in
the GRB afterglow emissions, e.g., produce a rebrightening as
observed in the afterglow of GRB 060206.

In our scenario illustrated in Fig. 1, the accretion-powered GRB
central engine firstly generates a jet (denoted by Jet $A$) along
LOS, which produces the observed prompt $\gamma$-ray emission. The
interaction between Jet $A$ and the ambient medium is responsible
for the first part of the afterglow until the large rebrightening. A
period $\Delta t$ later, the central engine ejects the second jet
(denoted by Jet $B$) at a larger viewing angle preventing
$\gamma$-ray detection. By assuming that the isotropic energy of Jet
$B$ is, however, significantly higher than that of Jet $A$, the
off-axis afterglow emission from Jet $B$ produces significant
rebrightening at lower energies. The two jets should also not
intersect, and therefore collide, with each other.

{\bf{Pre- Large Rebrightening.}} Emission from the forward shock,
driven by Jet $A$, interacts with the ambient medium to produce an
afterglow for up to 3000 s; during this time, the temporal decay
indices of the two small bumps are similar to the values predicted
by the standard fireball model. There is no signature of reverse
shock emission, which possibly ceases at very early times.
Alternatively, the reverse shock emission could be suppressed by the
magnetization of the ejecta (Zhang \& Kobayashi 2005) and many other
physical processes (Kobayashi 2000; Nakar \& Piran 2004; Kobayashi
et al. 2007; Jin et al. 2007). We consider that the cold Jet $A$
with a total isotropic kinetic energy
$E^{A}_{\rm{iso}}=10^{52}E_{52}$ erg propagates into the ambient
medium with a constant density $n_1$. A forward shock emerges and
energizes the surrounding materials by converting the bulk kinetic
energy of the jet into the internal energy of the shocked materials.
This internal energy is assumed to be shared by electrons and
magnetic fields with energy equipartition factors $\epsilon^A_e$ and
$\epsilon^A_B$, respectively. If the shock is adiabatic, the
synchrotron emission produced by slow cooling electrons is expected
to have a peak flux \beq F_{\nu,\rm max}=1.1\times 10^5
{\epsilon^A_B}^{1/2}E_{52}n_1^{1/2}D_{28}^{-2}(1+z)$ $\rm{\mu
Jy}\label{Maxflux} \eeq at time \beq
t_m=0.69{\epsilon^A_B}^{1/3}{\epsilon^A_e}^{4/3}E_{52}^{1/3}
\nu_{15}^{-2/3}(1+z)^{1/3}~ {\rm day}\label{maxtime} \eeq which is
defined to be the time at which the typical synchrotron frequency
$\nu_m$ crosses the observed frequency $\nu_{obs}=10^{15}\nu_{15}$
Hz (Sari et al. 1998). The luminosity distance of the burst is
$D_L=10^{28}D_{28}~{\rm cm^{2}}$.

For GRB 060206, no early flux peaks were detected in the optical and X-ray
bands, even at $t\sim200$ s. It is reasonable to
believe that the flux peak time was less than ${\sim 0.002~ {\rm
day} }$. A lower limit to the peak flux $F_{\nu,\rm max}\sim 1000{\rm
~\mu Jy}$ can be obtained accordingly, by
 extrapolating the R-band flux of the first post-bump back to $200~$s.
Since the optical flux peak was at time $t_m$ and a cooling break in
the X-ray afterglow was not observed, the order of the frequencies,
after the flux peak, was $\nu_m<\nu_{obs}<\nu_c<\nu_X$, where
$\nu_c$ is the cooling frequency and $\nu_X$ is the typical X-ray
frequency. The redshift of GRB 060206 is known, so the isotropic
energy of its first jet $E^{A}_{\rm{iso}}$ is $\sim 5.8\times
10^{52}$ erg.  The energy equipartition factors of magnetic fields
and electrons, according to Eqs. (\ref{Maxflux}) and
(\ref{maxtime}), can be constrained if the number density of the
ambient medium is known. The spectral analysis revealed a high
number density in the environment of GRB 060206 (Fynbo et al. 2006).
Given a rational value of $n_1=50$ $\rm{cm^{-3}}$, it can be proven
that $\epsilon^A_B\geq3\times10^{-5}$ and $\epsilon^A_e\leq0.05$.
Our subsequent numerical results are consistent with this
constraint. We calculate the forward shock synchrotron emission from
Jet $A$ that reproduces the first post-bump segment of the R-band
light curve of temporal index $\alpha_2=1.2\pm 0.5$, which implies
an electron energy spectral index $p=4\alpha_2/3+1=2.6\pm1.6$.

{\bf{Large Rebrightening.}} The large rebrightening is due to the
off-axis emission from the second beamed jet. The off-axis effects
can generate a fast rise light curve, and the gamma-ray emission,
from internal shocks in the second jet, would not trigger
$Swift$-\textbf{BAT} due to the large viewing angle and the
initially large Lorentz factor when the condition
$\Delta\theta_B-\theta_B>1/\Gamma_B$ is fulfilled, where $\Gamma_B$
is the initial bulk Lorentz factor of Jet B. The half opening angle
of Jet B can be estimated by the measured jet break time $t_b=53000$
s (e.g., Frail et al. 2001)
\begin{eqnarray}
  \theta_B&\sim&0.057\Big(\frac{t_b}{1~\rm{day}}\Big)^{3/8}
\Big(\frac{1+z}{2}\Big)^{-3/8} \Big(\frac{E^{B}_{\rm{iso}}}{5\times
10^{53}~
\rm{erg}}\Big)^{-1/8}\nonumber\\
&&\Big(\frac{n_1}{0.1 ~\rm{cm^{-3}}}\Big)^{1/8}.
\end{eqnarray}
The half opening angle and isotropic kinetic energy
$E^{B}_{\rm iso}$ of Jet
B can be obtained only by numerical fitting to the rebrightening
and the late afterglow light curve.

\section{Numerical Method}
When Jet $A/B$ sweeps up the ambient medium,
a pair of shocks could be generated, including a
forward shock propagating into the medium and a
reverse shock penetrating into the ejecta. However, as analyzed
above, the reverse shock driven by Jet $A$ can
be ignored for the afterglow phase of interest, whereas it
may be necessary to consider the contribution of
the reverse shock to the afterglow emissions for Jet
$B$. We would therefore describe the dynamics of the two jets in
different ways.
\subsection{Dynamics}
\subsubsection{Jet $A$}$\hspace{2mm}$ The evolution
of the forward shock driven by Jet $A$ with an
initial Lorentz factor $\Gamma_A$ follows the
equation (Huang et al. 2000) \beq
\frac{d\gamma^A_2}{dR_A}=--2\pi(1-{\rm cos}\theta_A)
{R_A}^2 n_1
m_p\frac{{\gamma^A_2}^2-1}{\varepsilon^A_2m^A_2+
2(1-\varepsilon^A_2)\gamma^A_2 m^A_2}, \label{dynA}\eeq where
$m_p$ is the mass of the proton, $\gamma^A_2$ is the
bulk Lorentz factor of the forward shock, and
$\varepsilon^A_2$ is the radiative efficiency of the
forward-shocked electrons. In our calculations, we take
adiabatic shock assumption and thus $\varepsilon_2^A=0$.
The swept-up mass $m^A_2$ by the shock is determined by
\beq \frac{dm^A_2}{dR_A}=2\pi {R_A}^2(1-{\rm cos} \theta_A )
n_1 m_p. \eeq Finally, for describing the temporal
behavior of the shock, we refer to the equation
\beq
\frac{dR_A}{dt}=\frac{\beta^A_2}{1-\beta^A_2}\frac{c}{1+z}.
\eeq

\subsubsection{Jet $B$}
$\hspace{2mm}$ With respect to Jet $A$,
the consideration of Jet $B$ is more complicated
because of the involvement of the reverse shock. The system is
divided into four regions by the two shocks and the
contact discontinuity surface: the unshocked medium, the
shocked medium, the shocked ejecta, and the unshocked
ejecta, which is denoted by 1-4, respectively. Since the
Lorentz factor $\gamma$ and energy density $e$ are
continuous along the contact discontinuity, we have
$\gamma^B_2=\gamma^B_3$ and $e^B_2=e^B_3$ (Sari \& Piran
1995). The hydrodynamics of the forward-reverse shock
pairs are determined by (Huang et al. 2000; Yan et al. 2007)
\beq
\frac{d\gamma^B_2}{dR_B}=-2\pi {R_B}^2 (1-{\rm
  cos}\theta_B)\frac{Q}{P},
\eeq
where the convenient parameters $Q$
and $P$
are defined in Yan et al. (2007) to be
\begin{eqnarray}
Q&=&({\gamma^B_2}^2-1)n_1m_p+(\gamma^B_2\gamma^B_{34}-\gamma^B_4)
(\gamma^B_4n_4m_p) (\beta^B_4-\beta^B_{RS})
\nonumber\\
P&=&m^B_2+m^B_3+(1-\varepsilon^B_2)(2\gamma^B_2-1)m^B_2+(1-\varepsilon^B_3)
(\gamma^B_{34}-1)m^B_3\nonumber\\
&&+(1-\varepsilon^B_3)\gamma^B_2m^B_3(\gamma^B_4-\frac{\gamma^B_4\beta^B_4}{\beta^B_2}),
\end{eqnarray}
 where $\beta^B_2$, $\beta^B_4$ and \beq
\beta^B_{RS}=\frac{\gamma^B_3\beta^B_3n^B_3-\gamma^B_4\beta^B_4n^B_4}
{\gamma^B_3n^B_3- \gamma^B_4n^B_4} \eeq are
the velocities of the shocked medium,
unshocked ejecta, and the reverse shock in the observer's frame,
respectively.
The relative Lorentz factor of region 3 with respect to
region 4 is $\gamma_{34}^B$. The radiative
efficiency $\varepsilon^B_i$ of region $i$ ($i$=2, 3) is
equal to zero for the adiabatic shock. The mass $m^B_2$
swept up by the forward shock
and $m^B_3$ by the reverse shock, can be calculated
respectively, by
\beq \frac{dm^B_2}{dR_B}=2\pi {R_B}^2(1-{\rm cos} \theta_B )
n_1 m_p,
\eeq
\beq
\frac{dm^B_3}{dR_B}=2\pi
{R_B}^2(1-{\rm cos} \theta_B)(\beta^B_4-\beta^B_{RS})
\gamma^B_4n^B_4m_p,
\eeq
where the comoving number density
of the unshocked ejecta is expressed
as $n^B_4=E^B_{\rm iso}/(4\pi {R_B}^2 \gamma^B_4\Delta m_p
c^2 )$ where $E^B_{\rm iso}$ and $\Delta$ are the
kinetic energy and the width of Jet $B$, respectively.
The temporal evolution of the radius $R_B$ of
Jet $B$ satisfies the same equation as that of Jet $A$, which reads
\beq
\frac{dR_B}{dt}=\frac{\beta^B_2}{1-\beta^B_2}\frac{c}{1+z}.
\eeq

After the reverse shock crosses the ejecta, the Lorentz
factor of
the reverse-shocked ejecta follows
$\gamma^B_3\propto {R_B}^{-7/2}$ and the mass of the shocked
ejecta remains constant $m^B_3=M^B_{ej}=E^B_{\rm iso}/(\Gamma_Bc^2)$. However, the forward shock
continues to sweep up the external medium, evolving in a
similar way to Jet $A$
described by Eq. (\ref{dynA}) as
\beq
\frac{d\gamma^B_2}{dR_B}=-2\pi(1-{\rm cos}\theta_B)
{R_B}^2 n_1 m_p\frac{{\gamma^B_2}^2-1}
{\varepsilon^B_2m^B_2+2(1-\varepsilon^B_2)\gamma^B_2m^B_2}.
\eeq

\subsection{Synchrotron Emission}
Due to the existence of the shocks, the bulk kinetic energy
of the ejecta should be gradually converted to the
internal energy $e$ of the shocked materials,
which is shared between magnetic fields, electrons and
protons according to the fractions $\epsilon_B$, $\epsilon_e$ and
$1-\epsilon_e-\epsilon_B$, respectively.

It is accepted in general that the synchrotron radiation of the
shocked electrons produces the observed X-ray and optical afterglow
emissions. In a detailed calculation, we consider, as often assumed,
that the electrons without energy losses are accelerated by the
shocks in a way described by a power law distribution
\begin{eqnarray}
{dN_e\over d\gamma_e} \propto \gamma_e^{-p},~~ \gamma_m<\gamma_e<\gamma_M
\end{eqnarray}
 with a minimum Lorentz factor
$\gamma_{m}$ and a maximum Lorentz factor $\gamma_{M}$. Another
critical Lorentz factor $\gamma_c$, above which the energy losses of
the electrons due to both synchrotron and inverse Compton (IC)
radiation is significant, was derived by Sari et al. (1998). The
actual distribution of the electrons should be given according to
the following cases (Sari et al. 1998; Yan et al. 2007) as
\begin{eqnarray}
\frac{dN_e}{d\gamma_{e}}&\propto& \cases{{\gamma_{e}}^{-2},~~~~~~~{\gamma_{c}<\gamma_{e}<\gamma_{m}} \cr
{\gamma_{e}}^{-(p+1)},~\gamma_{m}<\gamma_{e}<\gamma_{M}},~~\rm fast~ cooling,
\end{eqnarray}
\begin{eqnarray}
\frac{dN_{e}}{d\gamma_{e}}&\propto& \cases{{\gamma_{e}}^{-p},~~~~~~{\gamma_{m}<\gamma_{e}<\gamma_{c}} \cr
{\gamma_{e}}^{-(p+1)},~~~\gamma_{c}<\gamma_{e}<\gamma_{M}},~~\rm slow~ cooling.
\end{eqnarray}
To be specific, the three characteristic Lorentz factors for the
distribution of the electrons can be calculated by
$\gamma_{m}\approx\epsilon_{e}(m_p/m_e)(p-2)/(p-1) \gamma_{\rm
rel}$, $\gamma_{M}\approx 10^8(B/1{\rm G)}^{-1/2}/(1+Y)$ and
$\gamma_c=6\pi m_ec(1+z)/[(1+Y)\sigma_T\gamma{B}^2t]$, where the
magnetic field strength $B=\sqrt{8\pi\epsilon_{B}e}$, the relative
Lorentz factor $\gamma_{\rm rel}$ is taken to be $\gamma_{34}$ for
the reverse-shocked region and $\gamma_2$ for the forward-shocked
region, and the Compton parameter $Y$ is defined as the ratio
between the IC and synchrotron luminosities (Sari et al. 1998; Sari
\& Esin 2001).

Using the derived electron distribution,
we can calculate the synchrotron emissivity to be (Rybicki \&
Lightman 1979)
\begin{eqnarray}
\varepsilon'(\nu')={ \sqrt{3}q_e^3B\over m_ec^2}\int d\gamma_e {dN_e\over
d\gamma_e}f\left({\nu'\over\nu_*}\right),\label{epsilon}
\end{eqnarray}
where $q_e$ is the electron charge,
$\nu_*=3\gamma_e^2q_eB/(4\pi m_ec)$, $f(x)=x\int_x^{\infty}K_{5/3}(k)dk$
with $K_{5/3}(k)$ being the modified Bessel function.
By integrating over all of the emitting regions, we derive the observed
synchrotron flux density at a frequency $\nu$ to be (Huang et al. 2000)
\begin{eqnarray}
F_{\nu}(t)={1\over 4\pi D_{L}^2}\int dV'{\varepsilon'[\gamma(1-\beta\mu)\nu(1+z)]\over[\gamma(1-\beta\mu)]^3},
\end{eqnarray}
where $\mu=\cos\Theta$ is the cosine value of the angle $\Theta$
between the velocity of emitting materials and LOS. Taking into
account the time delay between the emissions from different
latitudes, the above integration should be performed on the
so-called equal-arrival-time surface that is determined by (Huang et
al. 2000) \beq t=\int \frac{1-\beta\mu}{\beta c}dR\equiv const. \eeq
From the obtained numerical synchrotron spectra, two break
frequencies appear, that is,
$\nu_{m}\simeq\gamma{\gamma_{m}}^2q_eB/2\pi m_ec$ and
$\nu_{c}\simeq\gamma{\gamma_{c}}^2q_eB/2\pi m_ec$, which correspond
to the characteristic Lorentz factors $\gamma_m$ and $\gamma_c$,
respectively. The peak flux at min\{$\nu_{m},\nu_{c}$\} is $F_{\rm
max}=\sqrt{3}\Phi q_e^3BN_{tot}/m_ec^2$, where $\Phi$ is an
integrating coefficient from Eq. (\ref{epsilon}) (Wijers \& Galama
1999). These characteristic quantities are applied in our analysis
in Sect 3.

We also take into account the
synchrotron self-absorption effect, which implies that a
correction should be applied to spectra below the synchrotron self-absorption
frequency $\nu_a$ as performed by Wu et al. (2003) and Zou
et al. (2005).

\subsection{Fitting the afterglow data}
Using the model described above with parameter values as listed in
Table 1, we describe numerically the R-band afterglow data of GRB
060206 in Fig. 2. As can be seen, the first post-bump in the light
curve can be reproduced well using the forward shock emission from
Jet $A$. Both the forward- and reverse-shock light curves of Jet $B$
are also presented in Fig. 2. The large rebrightening can be
attributed mainly to the forward shock emission from Jet $B$ because
the off-axis effect suppresses the peak flux of the reverse-shock
emission, which usually peaks at a few hundred seconds from its
beginning. For Jet $A$, we observe that the values of
$\epsilon^A_e$, $\epsilon^A_B$, and $n_1$ are consistent with our
analysis in Sect 3. For Jet $B$, we adopt typical values of 0.1 for
$\epsilon^B_e$ and 0.01 for $\epsilon^B_B$ since there is no
observational constraint on the shock parameters. The shock
parameters $(\epsilon_e, \epsilon_B, p)$ for different jets and/or
for different shocked regions may be different as found for the
two-component jets model (Jin et al. 2007) and the forward-reverse
shock model (Fan et al. 2002). The off-axis effect alone cannot
explain the fast rise in the large rebrightening. In addition, the
zero time effect can steepen the rise further (Zhang et al 2006;
Liang et al. 2006). We find that the time delay between the two jets
is $\Delta t=2000$ s, which agrees with the rising segment, the
subsequent normal decay phase, and the break in the light curve at
late times.

It should be noted that the parameters ($\theta_A$,
$\Delta\theta_A$, $\Gamma_A$) for Jet $A$, ($\theta_B$, $\Delta\theta_B$,
$\epsilon^B_e$, $\epsilon^B_B$, $\Gamma_B$, $E^B_{\rm iso}$, $\Delta$)
for Jet $B$ and the number density of the
  ambient medium $n_1$ are not exclusively
determined. The jet break time for Jet $A$ is hidden by the
dominating emission of Jet $B$, while the isotropic kinetic energy
of Jet $B$ cannot be measured because its gamma-ray emission cannot
trigger the detector. The most appropriate method for determining
the parameter values depends upon the following ingredients: (1) we
fit the observational R-band data of the first post-bump segment to
constrain the values of $\theta_A$ and $\Gamma_A$ combined with the
known isotropic gamma-ray energy of Jet $A$, (2) typical values of
$\epsilon^B_e$, and $\epsilon^B_B$ are used for Jet $B$, (3) the
off-axis angle $\Delta\theta_B$ cannot be too large, otherwise an
abnormally large isotropic energy for Jet $B$ is required, (4) the
condition of $\Delta\theta_B-\theta_B>1/\Gamma_B$ should be
satisfied to avoid the detection of gamma-ray emission from Jet $B$,
and (5) to simplify the calculation, we assume that the path of Jet
$A$ does not intersect with that of Jet $B$, which requires that
 $\theta_A-\Delta\theta_A\leq \Delta\theta_B-\theta_B$.

\section{Summary and Discussions}
We have presented a solution for the remarkable rebrightening
observed
in the afterglow of GRB 060206, which is
attributed to emission from an off-axis beamed jet
originating from the late activity of the central engine
after the prompt gamma-ray emission phase. We have argued that
the precession of the torus or accretion disk of the
central engine has caused the two jets to move in different directions.
Although we only attempt to describe the large rebrightening, it is
reasonable to speculate that the five bumps in the R-band
afterglow light curve of similar profiles may have the
similar origins: emission from different delayed off-axis
jets ejected by the GRB central engine.
The difference between the temporal indices of the five
post-bump
light curves indicates that the electrons in
different jets have different spectral indices $p$ for their
electron distributions. As a result, the spectra exhibit
an SED evolution, especially when the separate jets have
comparable contributions to the total emission. This
can be naturally explained by the significant SED evolution
in GRB 060206 detected by the Liverpool Telescope
(Monfardini et al. 2006).

The isotropic energy of the second jet is more than
one order of magnitude higher than the first jet in our
scenario. The collimation-corrected energy of Jet $B$
is $\sim 4.3$ times larger than Jet $A$, which is larger
than the energy required to explain the big rebrightening
in the energy injection model as mentioned in Sect 1.
The precise mechanism that triggers the central engine again
remains unknown. One possibility is that a mass of debris falls
back onto the central compact object, generating another
more energetic jet. Since our scenario can reproduce well
the observations of the large bump in GRB 060206, as a conservative
extrapolation, we propose that GRB 970508 and GRB 060210,
which display remarkable rebrightening, may
have in addition a precessing torus or accretion disk.

A further consequences of our scenario is that the jet break is
determined by the off-axis second jet rather than the first one,
which produces the main burst. In this case, we cannot measure the
isotropic energy of the second jet directly and must fit the
afterglow data to obtain an estimation. When a large bump appears in
GRB afterglows, we are therefore unlikely to be able to derive the
jet opening angle, using the break time and isotropic gamma-ray
energy release, because these two quantities originate in two
different jets (Stanek et al. 2007).

Finally, several models could be applied to explain afterglow
light curves exhibiting rebrightening. These include variable
external density profiles (Lazzati et al. 2002), refreshed shocks
(Granot et al. 2003; \Bj et al. 2004), and angular dependence of
the energy profile on the jet structure (Nakar et al. 2003), each
of which can play a role. Peculiar behavior in light
curves caused by the precession of the central engine was
discussed by Reynoso et al. (2006), although more observations are required
to identify its true nature.

\section{Acknowledgments}
We would like to thank the anonymous referee's
constructive comments and suggestions which improved our
paper significantly. X. W. Liu thanks D. M. Wei for his encouragement to accomplish
this work. We thank Y. F. Huang, Y. W. Yu, Y. Li and L. Shao for helpful
discussions. This work was supported by the National Natural
Science Foundation of China (grants 10473023, 10503012, 10621303,
and 10633040). XFW gratefully acknowledges Re'em Sari during his
visit to Caltech, also thanks the supports of China Postdoctoral
Science Foundation, K. C. Wong Education Foundation (Hong Kong),
and Postdoctoral Research Award of Jiangsu Province.

\begin{figure}[ht!]
  \includegraphics[angle=0,width=10cm]{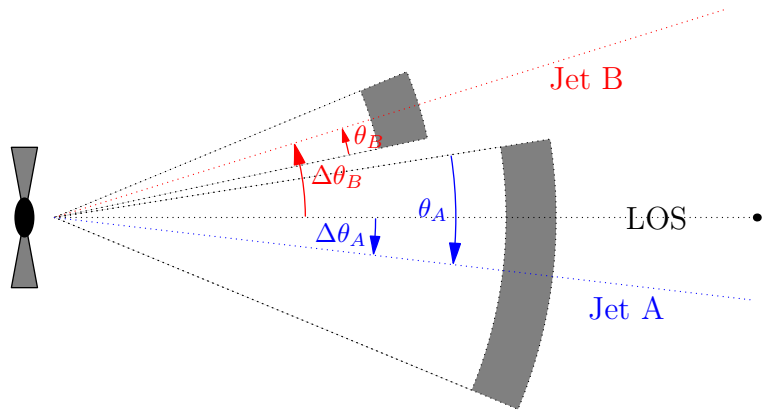}
  \caption{ Schematic two jets scenario for GRB 060206.}
  \label{fig1}
\end{figure}

\begin{figure}[ht!]
  \includegraphics[angle=0,width=10cm]{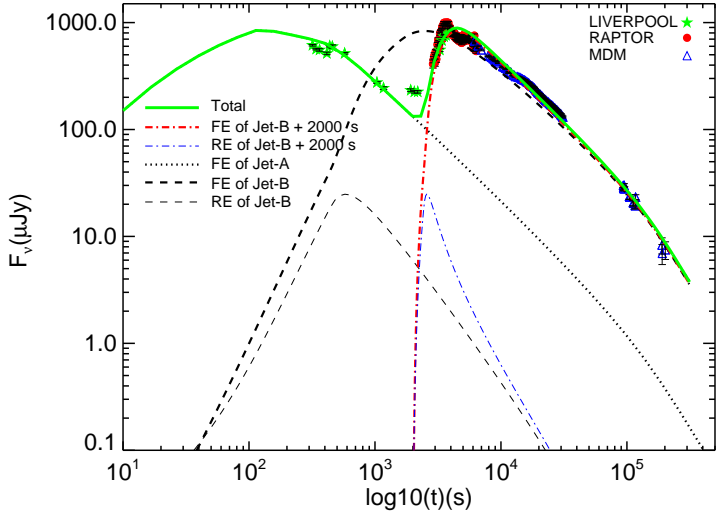}
  \caption{R-band light curve of
    GRB 060206. ``FE''
    and ``RE'' represent the forward shock emission and the reverse
    shock emission, respectively. The dotted line
    represents the contribution from Jet A. The thick, dashed
    line and the thin, dashed line correspond to the emissions from the
    forward shock and the reverse shock of Jet B,
    respectively. The dash-dotted lines represent the emission
    from Jet B taking into account the zero-time effect. The
    thick, solid line includes the contribution from both Jet A
    and Jet B. R-band data are taken from
    Monfardini et al. (2006), Stanek et al. (2007) and
    $\rm{Wo\acute{z}niak}$ et al. (2006).}
  \label{fig2}
\end{figure}

\newpage
\begin{table}{Table 1. The main parameters adopted in our calculations.\\ \\}
\begin{tabular}{cccccccccc}
\hline \hline
& &$\bold{Jet\: A}$       &\multicolumn{2}{c}{$\bold{Jet\: B}$} \cr
\hline
$\bold{Symbols}$ &$\bold{Definitions\: of\: the\: Symbols}$
                 &$\bold{Forward\: Shock}$ &$\bold{Forward\: Shock}$
                 &$\bold{Reverse\: Shock}$\cr

\hline
$E^l_{\rm iso}\:(\rm{erg})$  &$\rm{Isotropic\: Energy}$
                       &$5.8\times 10^{52}$
                       &\multicolumn{2}{c}{$1.0\times10^{54}$}\cr

\hline
$\Gamma_l$             &$\rm{Initial\:Lorentz\:Factor}$
                       &$300$
                       &\multicolumn{2}{c}{$300$}\cr

\hline
$n_1\:(\rm{cm^{-3}})$   &$\rm{Number\: Density\:of\: ISM}$
                       &$50.0$
                       &\multicolumn{2}{c}{$50.0$}\cr

\hline
$p_l$                    &$\rm{Electron\: Spectral\: Index}$
                       &$2.10$
                       &\multicolumn{2}{c}{$2.12$}\cr

\hline
$\theta_l$             &$\rm{Half\:Opening\:Angle}$
                       &$0.08$
                       &\multicolumn{2}{c}{0.04}\cr

\hline
$\Delta\theta_l$       &$\rm{Viewing\:Angle}$
                       &$0.07$
                       &\multicolumn{2}{c}{0.05}\cr
\hline
$\varepsilon^l_i$        &$\rm{Radiative\: Efficiency}$
                       &$0.0$
                       &$0.0$
                       &$0.0$\cr

\hline
$\epsilon^l_{e,i}$        &$\rm{Electron\:Equipartition\:Factor}$
                       &$0.05$
                       &$0.1$
                       &$0.1$\cr

\hline
$\epsilon^l_{B,i}$        &$\rm{Magnetic\:Equipartition\:Factor}$
                       &$0.0008$
                       &$0.01$
                       &$0.01$\cr
\hline
\hline
&\multicolumn{4}{c}{$\bold{Values\:Of\:Following\: Physical\:Quantities\:At\:t\sim3000\:s}$}\cr
\hline
\hline
$\gamma^l_i$             &$\rm{Bulk\:Lorentz\:Factor}$
                       &$22$
                       &$29$
                       &$32$\cr

\hline
$\gamma^l_{m,i}$          &$\rm{Minimal\:Lorentz\:Factor}$
                       &$180$
                       &$540$
                       &$3.5$\cr
\hline
$\gamma^l_{c,i}$          &$\rm{Cooling\:Lorentz\:Factor}$
                       &$2600$
                       &$205$
                       &$3.5$\cr

\hline
$\gamma^l_{M,i}$          &$\rm{Maximal\:Lorentz\:Factor}$
                       &$7.6\times 10^{7}$
                       &$3.6\times 10^{7}$
                       &$4.8\times 10^{7}$\cr


\hline
$B^l_i\: (\rm{G})$       &$\rm{Comoving\:Magnetic\:Field\:Strength}$
                       &$1.7$
                       &$7.8$
                       &$4.3$\cr

\hline
$m^l_i\: (\rm{g})$       &$\rm{Shocked\:Mass}$
                       &$1.2\times 10^{29}$
                       &$5.2\times 10^{29}$
                       &$3.7\times 10^{30}$\cr

\hline
\hline
\end{tabular}
\end{table}
\noindent Notes --- For Jet $A$, $l=A$ and for Jet $B$,
$l=B$. The subscript $i=2$ represents forward-shocked
region and $i=3$ represents reverse-shocked region. In the
lower part of the Table, some physical quantities involved in the
calculations at $t\sim 3000$s are presented for illustration.

\end{document}